\documentclass[preprint,showpacs,preprintnumbers,amsmath,amssymb]{revtex4}

\usepackage[textsize=tiny,backgroundcolor=yellow]{todonotes}

\usepackage{graphicx}
\usepackage{epstopdf}
\usepackage{dcolumn}
\usepackage{bm}

\newcommand{\be}{\begin{equation}}
\newcommand{\ee}{\end{equation}}
\newcommand{\bea}{\begin{eqnarray}}
\newcommand{\eea}{\end{eqnarray}}

\def\cm{\textrm{cm}$^{-1}$}
\newcommand{\bra}[1]{\left\langle#1\right|}
\newcommand{\ket}[1]{\left|#1\right\rangle}
\begin{document}

\title{Laser-Driven Localization of Collective CO Vibrations in Metal-Carbonyl Complexes}

\author{Mateusz Lisaj}
\author{Oliver K\"uhn}
\email{oliver.kuehn@uni-rostock.de}
\affiliation{
Institut f\"{u}r Physik, Universit\"{a}t Rostock, Universit\"{a}tsplatz 3, 18055 Rostock, Germany
}%

\date{\today}

\begin{abstract}
Using the example of a cobalt dicarbonyl complex it is shown that two perpendicularly polarized IR laser pulses can be used to trigger an excitation of the  delocalized CO stretching modes, which corresponds to an alternating localization of the vibration within one CO bond. The switching time for localization in either of the two bonds is determined by the energy gap between the symmetric and asymmetric fundamental transition frequencies. The phase of the oscillation between the two local bond excitations can be tuned by the relative phase of the two pulses. The extend of control of bond localization is limited by the anharmonicity of the potential energy surfaces leading to wave packet dispersion. This prevents such a simple pulse scheme from being used for laser-driven bond breaking in the considered example. 
\end{abstract}

\maketitle

\section{Introduction}
\label{sec:intro}
Collective vibrations of equal functional groups are a frequent phenomenon in symmetric molecules. They lead to distinct features in infrared (IR) absorption and Raman spectra according to the respective selection rules. While traditional IR spectroscopy had been focused on identifying species or establishing the  correlation with structural motifs, ultrafast laser spectroscopy and pulse shaping techniques provide direct access to a broader range of the potential energy surface (PES), covering its anharmonicity as well as bond-breaking channels. 
For instance,  anharmonic couplings between collective normal modes can be studied in quite some detail using ultrafast  two-dimensional IR spectroscopy. Taking metal-carbonyl compounds as an example, anharmonic couplings  \cite{baiz09_1395} and spectral diffusion \cite{baiz09_2433}  have been studied for Mn$_{2}$(CO)$_{10}$. 

Tailoring laser pulses such as to selectively break bonds has been a goal of laser-assisted reaction control for many years \cite{rice01,SfbBook,shapiro11}. 
Although there is a considerable number of theoretical studies, experimental realizations are scarce. A pre-requisite for IR selective bond-breaking is mode-specific vibrational ladder climbing. Here, chirped pulse excitation was shown to enhance ladder climbing efficiency  since the pulse frequency is adjusted instantaneously to the decreasing vibrational level spacing \cite{chelkowski90_2355}. This has been demonstrated for a number of  relatively small molecules in Refs. \cite{arrivo95_247,maas98_75,witte03_2021,windhorn03_641}. 
With the  development of flexible mid-IR pulse shapers based on acousto-optic modulators \cite{strasfeld09_1} it became possible to control specific populations  beyond what can be achieved with simple linearly chirped pulses \cite{strasfeld07_038102}. This was demonstrated for the
collective CO vibrations in W(CO)$_6$. An extension, now including polarization shaping, was later shown to be able to discriminate between excitation of different carbonyl-stetching modes in MnBr(CO)$_5$ \cite{strasfeld09_105046}.
Phase shaping laser control  of coherent superposition states corresponding to two different CO vibrations was demonstrated by Ashihara et al. for an iridium dicarbonyl complex in solution \cite{ashihara13_05024}.
Pushing vibrational ladder climbing towards reaction control it was shown experimentally that rather high excitation levels can be reached such as to enable bond breaking mediated by intramolecular vibrational energy redistribution (IVR) \cite{witte03_2021} or by direct excitation of the reaction coordinate \cite{windhorn03_641}. 

Due to their relevance for many chemical and biological processes breaking metal-carbonyl bonds with laser light is a particular interesting target. 
Carboxymyoglobin or -hemoglobin are among those systems, which attracted most interest in this respect. Although there is only a single CO group, proposed mechanisms and challenges are characteristic for control of this type of bond motion. 
In particular current laser sources enable only for an efficient excitation of the CO vibration. Hence, ultrafast metal-carbonyl bond breaking can only occur as a consequence of anharmonic coupling between the metal-carbon and the carbon-oxygen bond. Joffre et al. \cite{ventalon04_13216} have been the first to demonstrate coherent vibrational ladder climbing in a protein, exciting the anharmonic CO oscillator up to the $\nu=6$ level. Even though the energy of this level ($\sim$12700 \cm) is well above the threshold for Fe-CO bond breaking ($\sim$ 6000-7500 \cm), no reaction products could be observed. In order to understand this behavior, an all Cartesian reaction surface model had been developed in Ref. \cite{kuhn05_48}. Based on this model and using the multi-configuration time-dependent Hartree (MCTDH) method \cite{beck00_1} wave packet simulations have been performed, mimicking CO stretch vibrational excitation above the Fe-CO bond dissociation threshold. Here, it was found that the wave packet stays localized in the  CO-stretch coordinate up to 1.5 ps, what points to lacking anharmonic coupling to the Fe-CO center of mass motion. The same conclusion was reached in \cite{kuhn09_329} based on the analysis of simulations taking into account the chirped pulse explicitly for the model of Ref. \cite{kuhn05_48}  and using MnBr(CO)$_5$ as an mimic in Ref. \cite{gollub07_369}.
Vibrational ladder climbing in carboxyhemoglobin was also addressed by Meier and co-workers. In Ref. \cite{meier05_044504} local control theory (for a review, see Ref. \cite{engel09_29}) was applied to a two-dimensional non-reactive model to show that upon extending the pulse duration, the predicted pulse shape changes from a simple chirp to a multiple sweep form, composed of several chirped pulses. This model was later supplemented by the effect of the fluctuating protein environment on the excitation efficiency of the CO-stretching mode \cite{falvo13_145101,debnath13_12884}. 
In passing we note that control theory has also been used in a proof of principle  study of metal-CO bond dissociation in HCo(CO)$_4$ by \emph{direct} excitation of the metal-carbonyl bond \cite{orel00_94}. Further studies on this molecule focused on the competitive Co-H vs. Co-CO bond breaking \cite{zhao99_7,kuhn99_3103,zhao00_4882}.

Viewed in the context of laser control of molecular dynamics, the collective nature of the CO normal mode vibrations in cases of multiple CO groups introduces even another challenge, i.e. the energy is not deposited into a {\it single} bond, but spread over {\it many} bonds within the molecule. This can be expected to make non-statistical bond-breaking impossible. Therefore, a pre-requisite for selective metal-carbonyl bond breaking in compounds with multiple CO groups would be a localization of the vibration along a single CO bond. Whether this can be achieved using femtosecond IR laser pulses is the main topic of this paper.  

In Sec. \ref{sec:methods} we first introduce the model system which is a cobalt dicarbonyl complex. In order to describe the symmetric and antisymmetric carbonyl stretching vibrations as well as possible metal-carbonyl bond breaking a four-dimensional (4D) model is developed. Respective potential (PES) and dipole moment surfaces (DMS) are calculated using density functional theory and the quantum dynamics in the laser field is solved numerically. The results Sec. \ref{sec:res} starts with a discussion of the potential energy surface and the IR absorption spectra. Their analysis suggests a scheme for the subsequently studied  laser-driven localization of vibrations. Finally, the possibility of bond breaking is considered, using wave packets that explore the PES starting from different initial conditions. Conclusions are presented in Sec. \ref{sec:conc}.
\section{Methods}
\label{sec:methods}
\subsection{Model System}
\begin{figure}[t]
    \centering
    \includegraphics[width=0.5\textwidth]{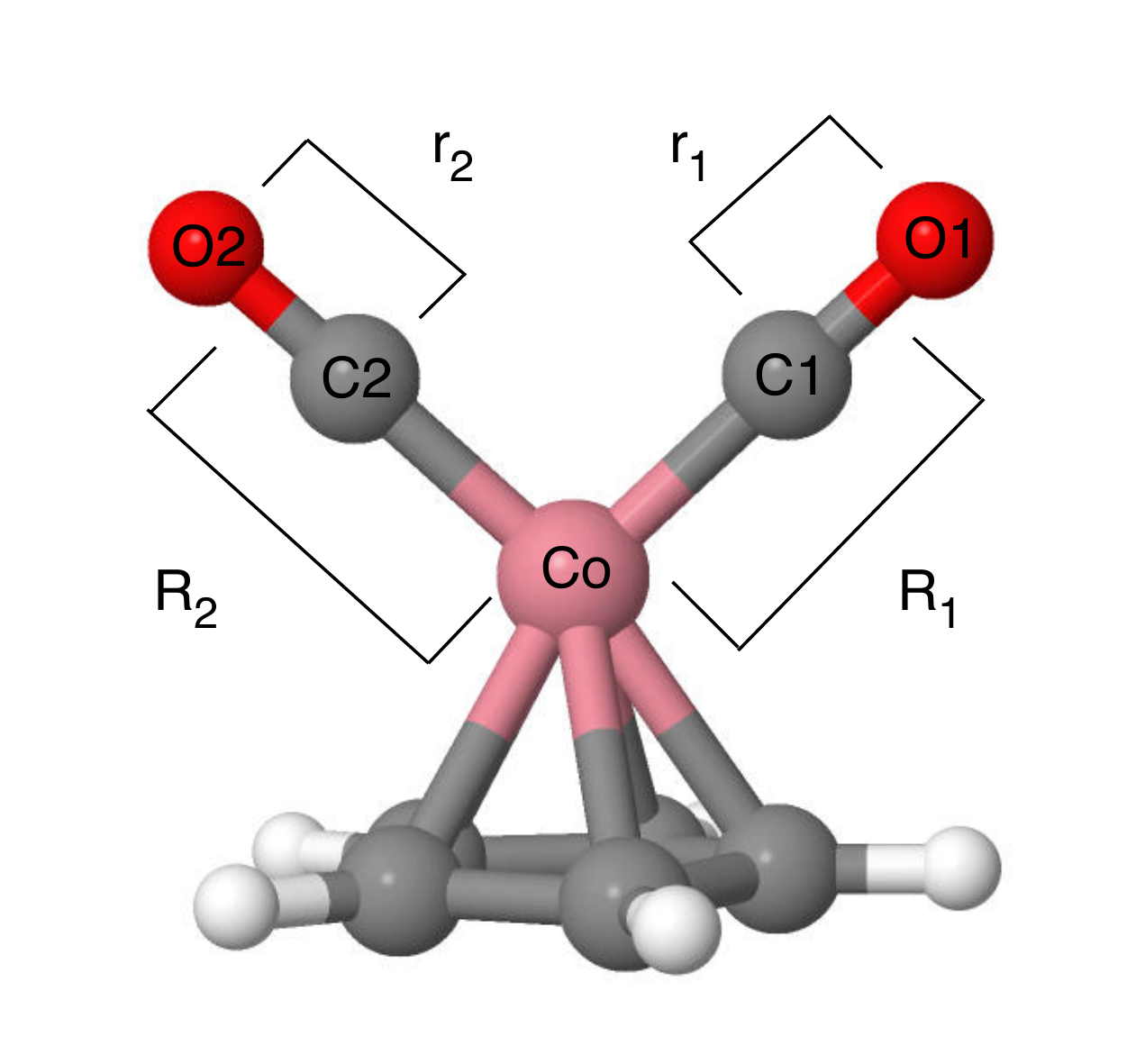}
    \caption{Equilibrium structure of CpCo(CO)$_{2}$ at the DFT/B3LYP (LanL2DZ) level of theory. The coordinates for the CO vibration ($r_{1,2}$)  and for Co-C bond dissociation ($R_{1,2}$) are defined along the two bond directions. Note that it is assumed that the Co(CO)$_2$ fragment is in the $X-Y$ laboratory frame with the $X$-axis bisecting the angle C1-Co-C2. The dipole moment is -2.78 D and oriented along the $X$-direction.}
    \label{fig:geo}
\end{figure}

In order to demonstrate collective mode localization we consider the dicarbonyl complex  CpCo(CO)$_{2}$ (cyclopentadienyl cobalt dicarbonyl, (C$_{5}$H$_{5}$)Co(CO)$_{2}$). This choice is motivated by the relatively low Co-CO bond dissociation energy of this complex. Quantum chemical calculations have been performed using density functional theory (DFT) with the B3LYP functional and the 
LanL2DZ basis set as implemented in the Gaussian 09 program package \cite{g09}. This level of theory had been tested, for instance, in Ref. \cite{dunietz03_5623} for the CO ligated heme group. 
The optimized ground state geometry is shown in Fig. \ref{fig:geo}. In accordance with electron diffraction data \cite{beagley79_47} the structure is eclipsed, i.e. the Co(CO)$_2$ plane passes through a ring carbon atom. Equilibrium bond lengths are found to be 1.75 \AA{} and 1.18 \AA{} for the Co-C(1,2) and C(1,2)-O(1,2) bond, respectively. This is in good agreement with electron diffraction data, which give 1.69 \AA{} and 1.19 \AA{} for the two bond lengths \cite{beagley79_47}.

In order to describe metal-carbonyl bond vibration and dissociation a four-dimensional 4D model has been adopted as shown in Fig. \ref{fig:geo}.
This includes the center-of-mass coordinates of the two CO groups with respect to the metal center, $R_1$ and R$_2$, 
as well as the CO bond coordinates $r_1$ and $r_2$. Below the coordinates will be given with respect to their equilibrium values mentioned above. The motion is assumed to take place along the two bond directions only. PES and DMS are calculated for a combination of the Fourier grid ($R_{1,2}$) and a grid given by the harmonic oscillator discrete variable representation ($r_{1,2}$). For the Fourier grid 80 points in the interval [-1.73:5.11] a$_{\rm B}$ have been used, whereas the bond motion is described by 17 points in the interval  [-0.50:1.00] a$_{\rm B}$. Hence there is a total of about 1.85 million grid points. Note that all coordinates, which are not explicitly considered are frozen at their equilibrium values.

\subsection{Quantum Dynamics}
\label{sec:qdyn}
The model Hamiltonian is given by 
\begin{eqnarray}
H(t)&=& H_{\rm mol} +H_{\rm field}(t) \nonumber\\
 &=& \sum_{i=1}^{2}\left(\frac{p_{i}^{2}}{2\mu_{\rm red}}+ \frac{P_{i}^{2}}{2M_{\rm CO}} \right) + V(r_{1},r_{2},R_{1},R_{2})
\nonumber\\
& -&\mathbf{d}(r_1,r_2,R_1,R_2) \cdot {\mathbf E}(t) \, .
\end{eqnarray}
Here, $ V(r_{1},r_{2},R_{1},R_{2}) $ and $ \mathbf{d}(r_1,r_2,R_1,R_2) $ are the PES and DMS, respectively, $ \mu_{\rm red} $ is the reduced mass of the CO vibration, and $M_{\rm CO}$ is the total CO mass.

The time-dependent Schr\"odinger equation 
\begin{equation}
i\hbar\partial_{t}\vert\Psi(t)\rangle=(H_{\rm mol}+H_{\rm field}(t))\vert\Psi(t)\rangle
\end{equation}
has been solved using the MCTDH method \cite{meyer90_73,beck00_1} as implemented in the Heidelberg program package \cite{mctdh84}.
On the grid specified above, the wave packet is represented by single particle functions (SPFs). Their number differs for the different simulations and will be specified below. Mode combination of $(r_1,r_2)$ and $(R_1,R_2)$ has been used. The integration has been performed within the constant mean field scheme and using a short-iterative Lanczos and a Bulirsch-Stoer integrator for the coefficients and the SPFs, respectively.
The PES and DMS have been fit to a product representation using the {\it potfit} module of the MCTDH package  \cite{jackle96_7974}. 

Besides expectation values of energies and coordinates the IR absorption spectrum will be calculated using \cite{may11}
\begin{equation}
\label{eq:spec}
A(\omega)=A_0 \omega \sum_{\alpha=x,y,z}{\rm Re}\int_{0}^{\infty} dt e^{i\omega t-\gamma t}\langle\Psi_{0}\vert d_{\alpha}U(t)d_{\alpha}\vert\Psi_{0}\rangle \, .
\end{equation}
Here, $A_0$ is a normalization constant, $\gamma$ is a parameter that mimics phenomenological broadening, $U(t)=\exp(-iH_{\rm mol}t/\hbar)$ the molecular time evolution operator, and $ \vert\Psi_{0}\rangle $ is the ground state, which is obtained by imaginary time propagation.

Further, the reaction yield for bond breaking will be studied. It is defined by means of a step-function operator, placed into the exit channel of the PES, i.e.
\begin{equation}
\label{eq:yield}
Y(t)= \bra{\Psi(t)} \Theta(R_1-R_{\rm exit})+\Theta(R_2-R_{\rm exit}) \ket{\Psi(t)}
\end{equation}
In Sec. \ref{sec:bondbreaking} the dividing surface is placed at $R_{\rm exit}=1.5 $ a$_{\rm B}$, cf. Fig. \ref{fig:fig_pes}a. Notice, that since no appreciable dissociation was observed absorbing boundary conditions have not been used.

\section{Results}
\label{sec:res}
\subsection{PES and DMS}
\begin{figure}[t]
    \centering
    \includegraphics[width=0.4\textwidth]{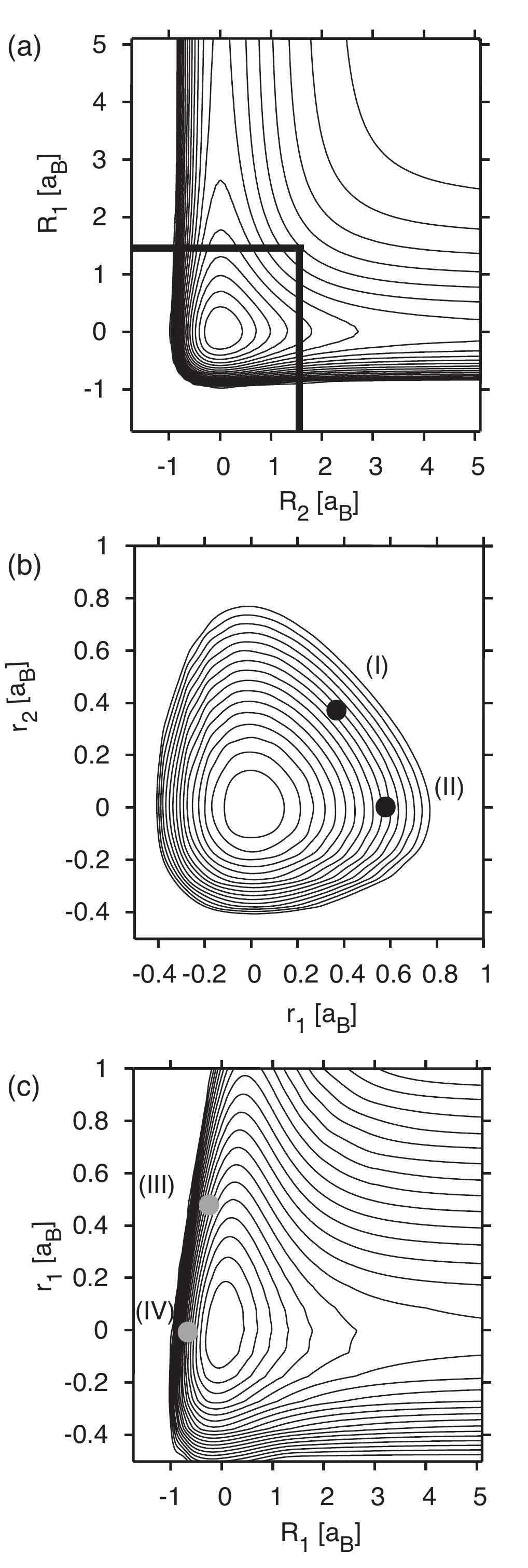}
    \caption{Cuts of the 4D PES with the other coordinates taken at their equilibrium values. The contour lines in (a) and (c) are from 3000 to 60000 \cm{} in steps of 3000 \cm. In panel (b) the contour  lines are from 2000 to 32000 \cm{} in steps of 2000 \cm. The solid lines in panel (a) mark the range beyond which the dissociation channels start. Points I-IV in panels (b,c) give the different initial displacement for the bond breaking study  in Sec. \ref{sec:bondbreaking}.}
    \label{fig:fig_pes}
\end{figure}

Before presenting results of the quantum dynamics simulations, PES and DMS will be  discussed. Various cuts of the PES are shown in Fig. \ref{fig:fig_pes}.
In panel (a) the PES along the two dissociation coordinates is shown. The dissociation energy at the end of the grid is  20 365 cm$^{-1}$ along $R_1$ and 20 191 cm$^{-1}$ along $R_2$. The slight difference comes from the fact that in the projection  the vector associated with $R_1$ points along the CH bond of the Cp ring, whereas for $R_2$ it bisects a CC bond. The same effect is responsible for the asymmetry with respect to $r_1=r_2$ of the PES for CO bonding vibrations shown in  Fig. \ref{fig:fig_pes}b. Overall, this figure also illustrates the anharmonicity, which is notable  already for the lowest contours. Finally, panel (c) gives the PES along the mixed bond vibration and bond dissociation coordinates. In the limit of large $R_1$ one clearly notices the expected CO bond compression. Further, there is a tilt of the PES around the equilibrium structure which points to the anharmonic coupling between these two coordinates. CO bond dissociation does not occur within the range of $r_1$.

\begin{figure}[t]
    \centering
    \includegraphics[width=0.4\textwidth]{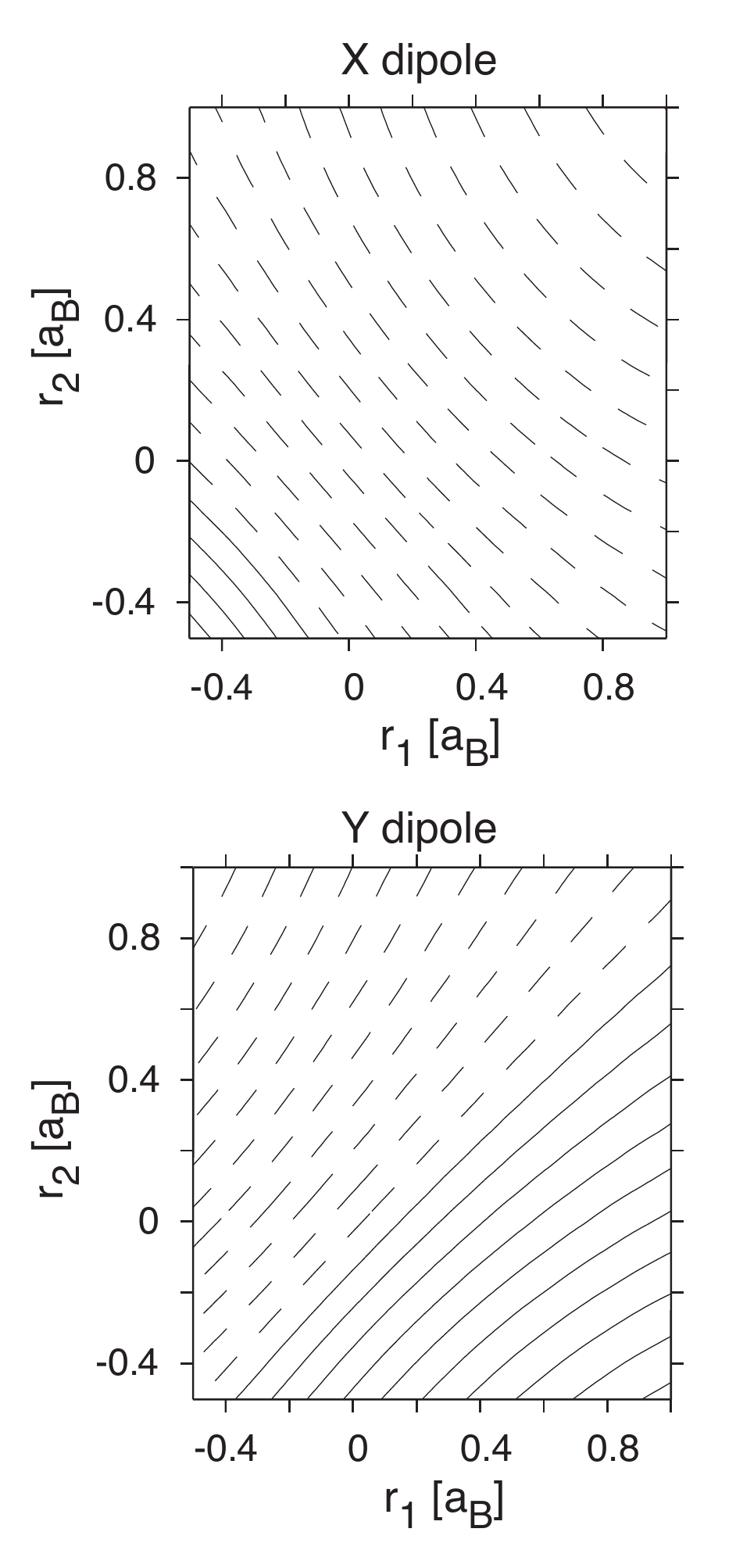}
    \caption{DMS along the bond vibration coordinates with $R_{1,2}$ being fixed at their equilibrium values. The upper and lower panel show the $X$ and $Y$ component of the dipole moment vector, respectively. The contours start from -3.2  ea$_{\rm B}$ and -2.15 ea$_{\rm B}$ in steps of 0.2 for the  $X$-  and $Y$- component, respectively.  The solid/dashed lines give positive/negative values. Note that the $Z$-component of the  dipole moment doesn't change along these coordinates.}
    \label{fig:fig_DMS}
\end{figure}

The DMS along the bond vibration coordinates to be used in the excitation schemes below is shown  in Fig. \ref{fig:fig_DMS}. The gradient, which is easily discernible from these figures, indicates that IR excitation will yield an $X$-polarized symmetric vibration ($\nu_{\rm s}$) along the collective coordinate $q_{\rm s}= (r_1+r_2)/2$ and an $Y$-polarized antisymmetric vibration ($\nu_{\rm a}$) along the collective coordinate $q_{\rm a}= (r_2-r_1)/2$.

\subsection{IR Spectrum}

\begin{figure}[t]
    \centering
    \includegraphics[width=0.6\textwidth]{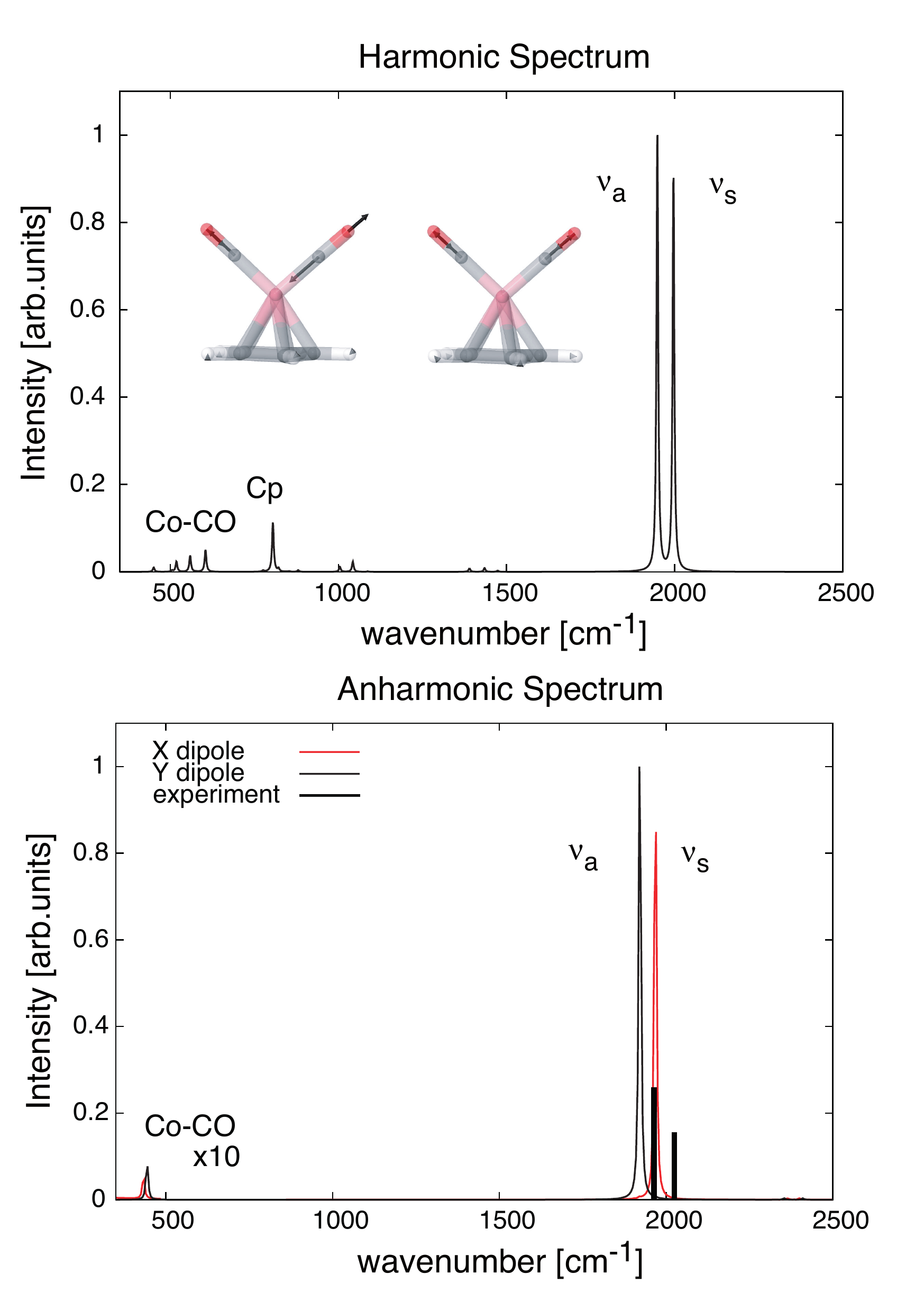}
    \caption{IR spectrum in harmonic approximation (upper panel, Lorentzian broadening of width 3 \cm) and including anharmonicity according to the present 4D model (lower panel, $\gamma$=1.667 ps$^{-1}$, total propagation time 8 ps). The range around 2000 \cm{} is shaped by symmetric and antisymmetric collective CO vibrations (normal mode displacements in insert of upper panel), which are polarized differently. The experimental data are taken from Ref. \cite{cotton55_175}}
    \label{fig:fig_IR}
\end{figure}

The harmonic and anharmonic IR spectra (cf. Eq. \eqref{eq:spec}) are given in Fig. \ref{fig:fig_IR}. The interesting region of the collective CO stretching vibrations is located around 2000 \cm. In harmonic approximation the antisymmetric and symmetric CO vibration are found at 1949 cm$^{-1}$ and 1997 cm$^{-1}$, respectively. Vibrations of the Cp ring are located around 800 \cm{} and Co-CO vibrations around 559 \cm. Upon including anharmonicity within the 4D model the CO vibrations change to 1922 \cm{} ($Y$-polarized $\nu_{\rm a}$) and 1968 \cm{} ($X$-polarized $\nu_{\rm s}$) , i.e. as expected their frequency decreases. Further, the model predicts two transitions at 433 and 443 \cm, which are due to the bond dissociation coordinate motion and will not  be considered further in the following. For the present gas phase model there are no experimental data available. IR spectra taken in CHCl$_3$ solution yield frequencies of 1967 \cm{} and 2028 \cm{} \cite{cotton55_175}.

\subsection{Localization of Collective CO Vibrations}
In the following we will assume that the molecule is fixed in the laboratory frame according to the model explained in Fig. \ref{fig:geo}. Since the $Z$-component of  dipole moment vector of the 4D model doesn't change, it suffices to consider the $X-Y$ plane only (unit vectors $\mathbf{e}_{X}$ and $\mathbf{e}_{Y}$). The electric field will be assumed to have the form
\begin{equation}
{\mathbf E}(t)={\mathcal{E}}(t) (\mathbf{e}_{X}\cos(\omega_{X}t + \phi_{X})+\mathbf{e}_{Y}\cos(\omega_{Y}t + \phi_{Y})) \,.
\label{eq:field}
\end{equation}
Here, ${\mathcal{E}}(t)$ is the field envelope given by
\begin{equation}
{\mathcal{E}}(t)=E_{0}\exp(-2\ln2(t-t_{0})^{2}/\tau^{2})\, ,
\end{equation}
where $E_{0}$ is the field amplitude and  $\tau$ is the full width at half maximum of the Gaussian that is centered at $t_0$. Further, Eq. \eqref{eq:field}
contains the carrier frequencies, $\omega_{X,Y}$, and the phases of the different polarization directions. In fact, only the relative phase, $\Delta \phi= \phi_{Y}-\phi_X$, will be of interest in the following.

\begin{figure}[t]
    \centering
    \includegraphics[width=0.6\textwidth]{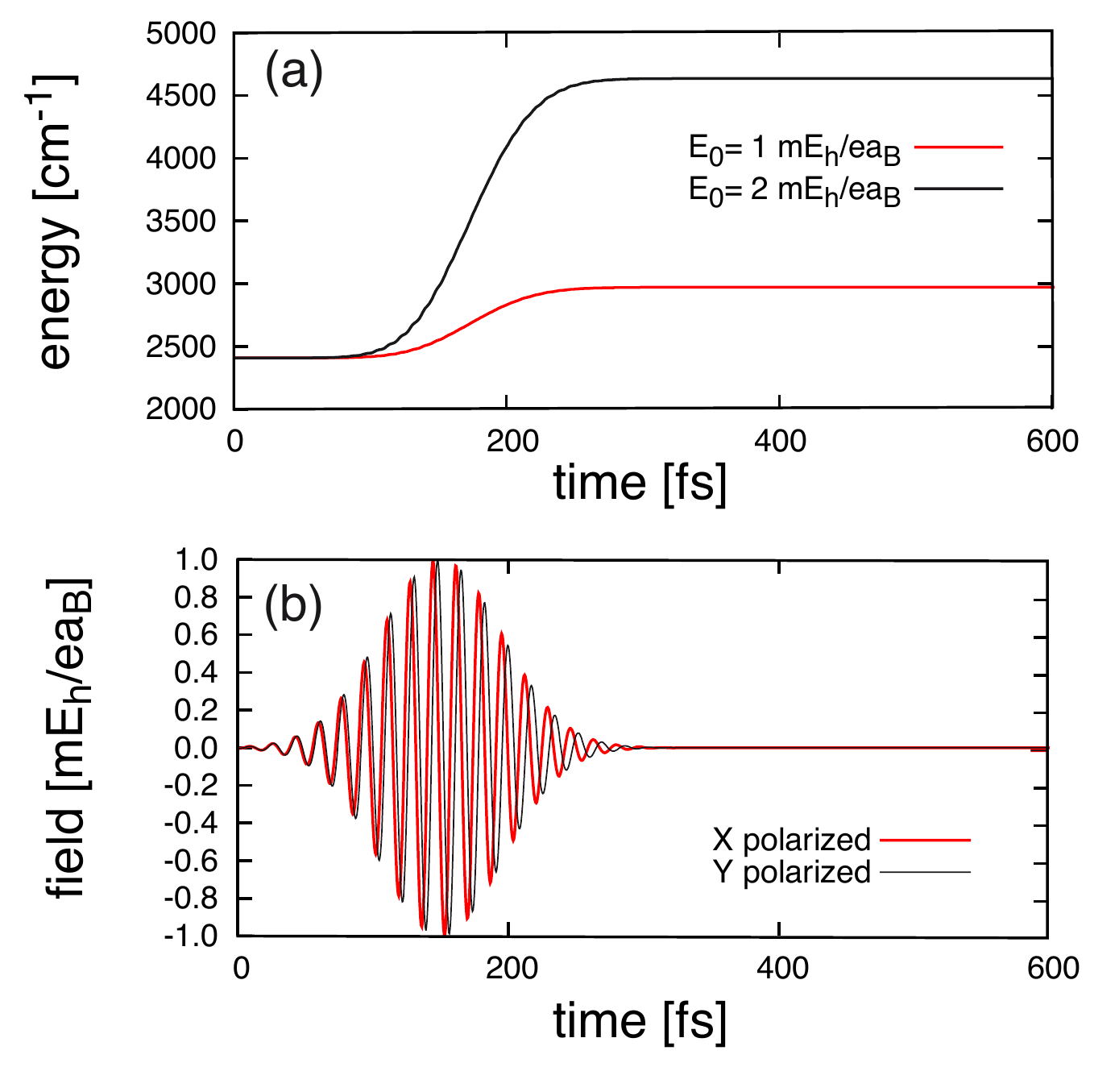}
    \caption{(a) Expectation value of the four-dimensional Hamiltonian for the laser fields given in panel (b). The pulse parameters are $ \omega_{X} = 1968$ \cm, $ \omega_{Y} =1922$ \cm, $t_0= 150$ fs, and $ \tau= 75$ fs. Two different field amplitudes are used, i.e. $E_0=1.0$ mE$_{\rm h}$/ea$_{\rm B}$ and $E_0=2.0$ mE$_{\rm h}$/ea$_{\rm B}$. The relative phase has been set to $\Delta \phi =0$ degrees; the energy expectation value is not appreciably influenced by the alternative choice of $\Delta \phi =180$ degrees used below.
    }
    \label{fig:fig_field}
\end{figure}

The frequencies have been chosen to be resonant to the fundamental transition of the $\nu_{\rm a}$ and $\nu_{\rm s}$ mode, polarized in $Y$ and $X$ direction, respectively. Two different  field amplitudes will be considered, each selected such that the expectation value of the 4D molecular Hamiltonian continuously rises up to a certain energy for the given duration $\tau$. The energy uptake of the molecule for the two cases is about 560  and 2230 \cm{} as shown in Fig. \ref{fig:fig_field}a; the pulses are given in panel (b) of that figure. In terms of the PES for bond motion in Fig. \ref{fig:fig_pes}b these energies are still in the part where the anharmonicity of the PES is modest. Further, these energies are well below the dissociation threshold, i.e. wave packet motion will occur  approximately in the range up to the second contour level in Fig. \ref{fig:fig_pes}c.

\begin{figure*}[t]
    \centering
    \includegraphics[width=1\textwidth]{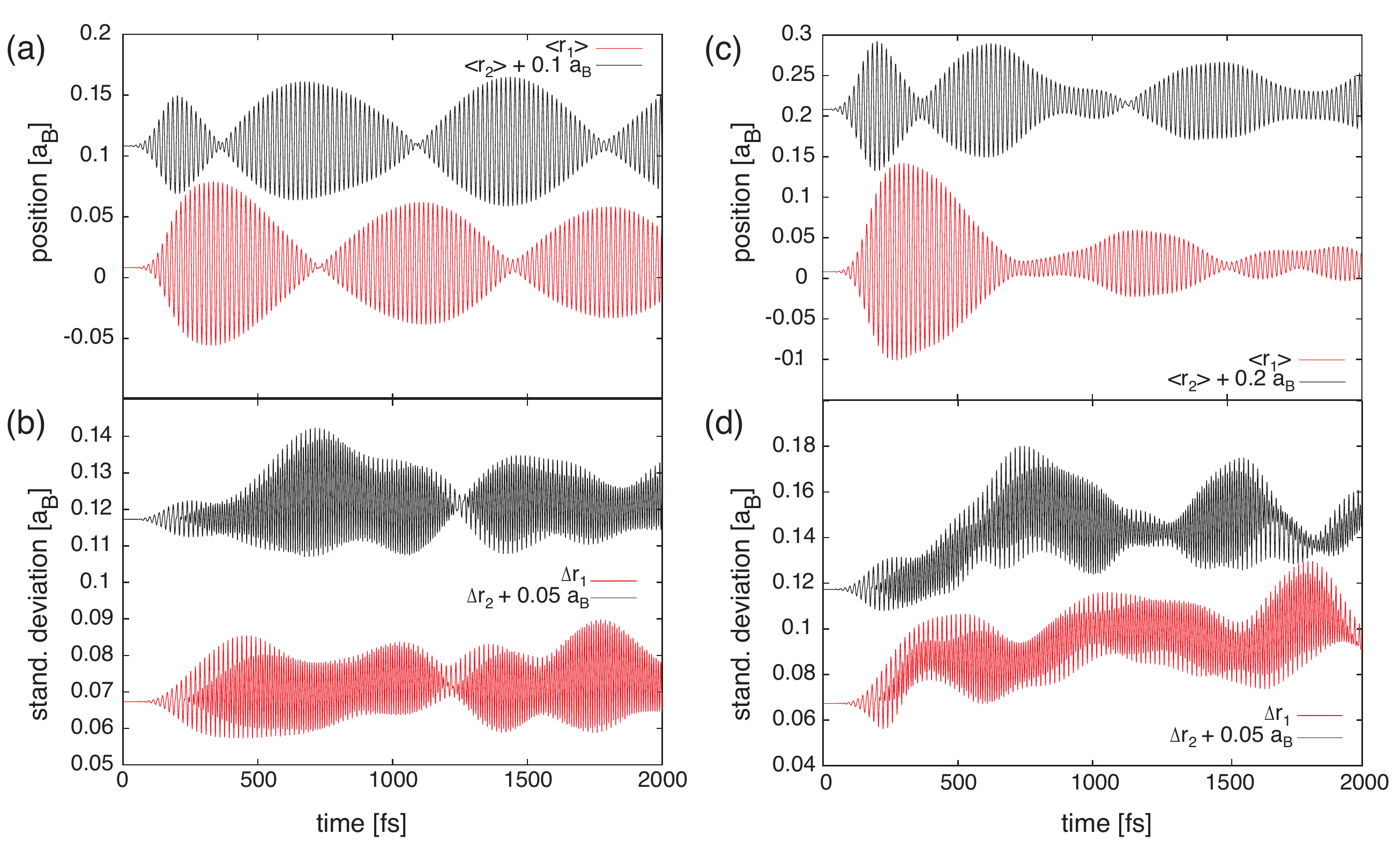}
    \caption{Coordinate expectation values and their standard deviation for CO bond coordinates and two different laser field amplitudes ((a,b): $E_0=1.0$ mE$_{\rm h}$/ea$_{\rm B}$, (c,d): $E_0=2.0$ mE$_{\rm h}$/ea$_{\rm B}$). The relative phase between the $X$ and $Y$ polarized fields is set to zero. For other field parameters, see Fig. \ref{fig:fig_field}.  The number of SPFs for the combined modes has been 4 and 12 in (a,d) and (c,d), respectively.
    }
    \label{fig:fig_pos_phase0}
\end{figure*}

In Fig. \ref{fig:fig_pos_phase0} the expectation values of the two bond distance coordinates are given for two different field amplitudes and relative phase $\Delta \phi = 0$. Panel (a) shows the result for a weak-field excitation, i.e. $E_0=1.0$ mE$_{\rm h}$/ea$_{\rm B}$. Clearly, the bond oscillations are modulated such that if $\langle r_1 \rangle$ has its minimum  $\langle r_2 \rangle$ has a maximum and vice versa. The period of this modulation is about 363 fs. In terms of the dynamics on the bond distance PES this implies that there are certain periods where the motion is essentially along $r_1$ or $r_2$ only. In other words, whereas each pulse separately would excited a collective vibration, i.e. along $q_{\rm a}$ or $q_{\rm s}$, a superposition of  two pulses leads to an alternating localization of the vibrational motion in the individual CO bonds. The wave packet associated with this  localized vibration, however, does not correspond to an eigenstate of the system. Similar to the well-know problem of tunneling in a double well potential, we observe a wave packet motion between two localized vibrations with a time scale for ``switching'' of  $ \pi /(\omega_{\rm s}-\omega_{\rm a}) = 363$ fs.
Starting around 1500 fs we notice that the localization in one coordinate is no longer perfect, i.e. the wave packet become increasingly influenced by the anharmonicity of the potential and the above relation no longer holds. During the propagation the wave packet stays rather compact, what can be seen from Fig.  \ref{fig:fig_pos_phase0}b where the standard deviation of the coordinate is plotted. Moreover, the expectation values of the dissociation coordinates stay almost constant (not shown). This indicates that anharmonicity is still rather modest under these excitation condition. 

Increasing the field amplitude to $E_0=2.0$ mE$_{\rm h}$/ea$_{\rm B}$ more energy is absorbed  by the molecular system, cf. Fig. \ref{fig:fig_field}a. This comes along with the exploration of the more anharmonic parts of the PES. As a consequence the wave packet dispersion is more pronounced and the simple two state superposition picture should break down. Nevertheless, the weak-field behavior  of alternating oscillations is still discernible from Fig. \ref{fig:fig_pos_phase0}c. Notice, however, that in particular the oscillation amplitude of $\langle r_1 \rangle$ decreases. Since the standard deviation increases at the same time (Fig. \ref{fig:fig_pos_phase0}d), the delocalized wave packet is no longer well described by the coordinate expectation value, i.e. the classical picture of a vibrating bond coordinate doesn't apply.

In order to rationalize this field-induced transient localization, we note that the 
molecule-field interaction can be written approximately as 
\begin{equation}
H_{\rm field}(t) \approx {\mathcal{E}(t)} (q_{\rm s} \cos(\omega t) + q_{\rm a} \cos(\omega t + \Delta \phi)) \, .
\end{equation}
Hence, for the choice of $\Delta \phi = 0$, the field effectively drives the coordinate $q_{\rm s}+q_{\rm a}=r_2$, i.e. the superposition which corresponds to localized motion along $r_2$, cf. Fig. \ref{fig:fig_pos_phase0}. For the case of the weaker field this corresponds to the excitation of a superposition of the fundamental transitions of the $\nu_{\rm a}$ and $\nu_{\rm s}$ modes, what explains the observed switching time.

\begin{figure*}[t]
    \centering
    \includegraphics[width=1\textwidth]{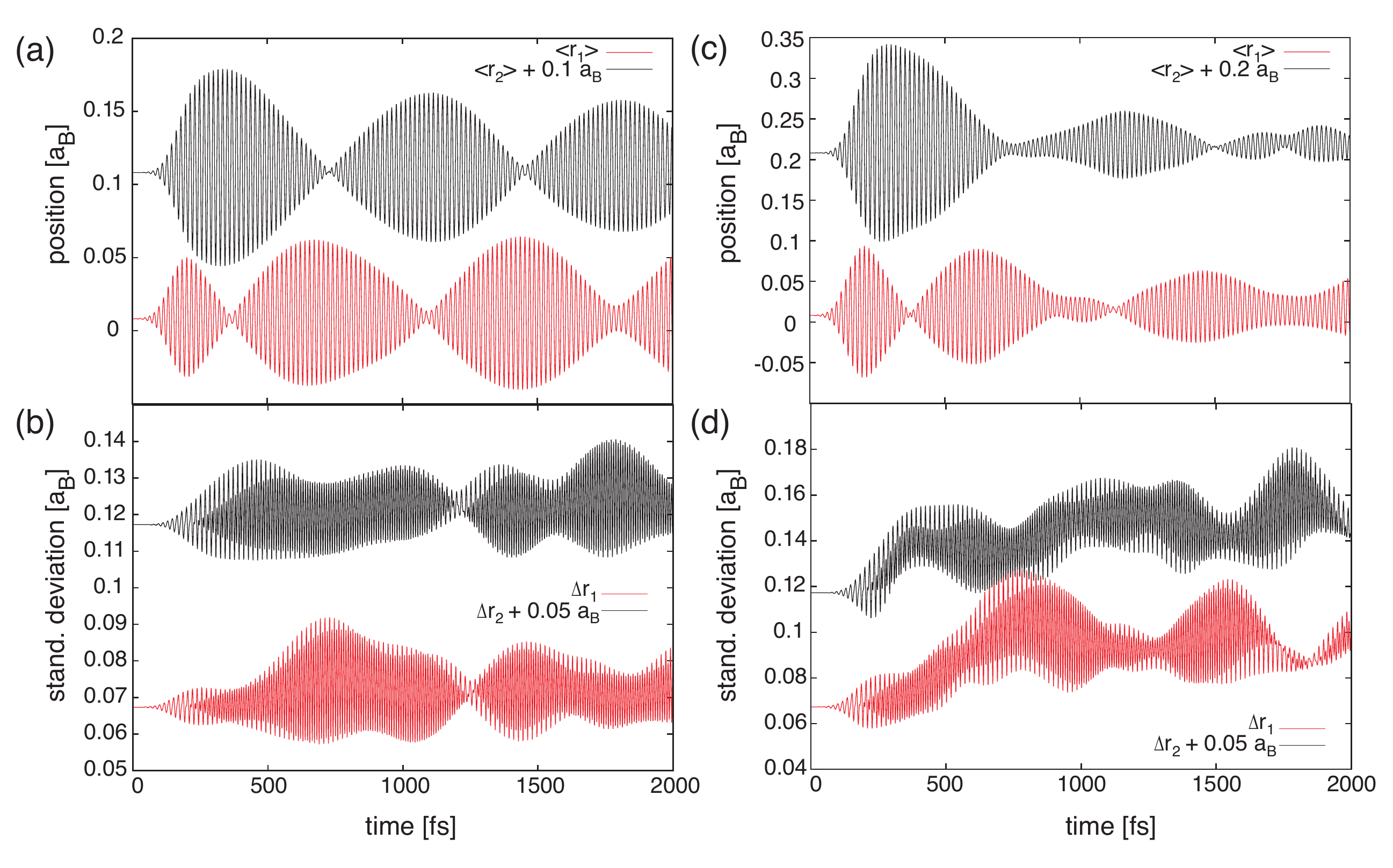}
    \caption{Coordinate expectation values and their standard deviationfor CO bond coordinates and two different laser field amplitudes ((a,b): $E_0=1.0$ mE$_{\rm h}$/ea$_{\rm B}$, (c,d):  $E_0=2.0$ mE$_{\rm h}$/ea$_{\rm B}$). The relative phase between the $X$ and $Y$ polarized fields is set to 180 degrees. For other field parameters, see Fig. \ref{fig:fig_field}. The number of SPFs for the combined modes has been 4 and 12 in (a,d) and (c,d), respectively.
    }
    \label{fig:fig_pos_phase180}
\end{figure*}

For the alternative choice of $\Delta \phi = 180$ degrees one would expect a driving of $q_{\rm s}-q_{\rm a}=r_1$. Indeed, this argument holds as can be seen from Fig. \ref{fig:fig_pos_phase180}a,b. Since there is not much difference in the PES concerning the two directions, a similar behavior is found upon increasing the field amplitude, cf. Fig. \ref{fig:fig_pos_phase180}c,d. 

Finally, we note that the broad pulse spectra leave some flexibility concerning the actual carrier frequencies. Exemplary, this has been investigated for the case where the carrier frequencies are both tuned mid-way between the $\omega_{\rm a}$ and  $\omega_{\rm s}$ transitions. As shown in the Supplementary Material this doesn't not result in an appreciably change of the bond oscillation pattern.
 
\subsection{Feasibility of Bond Breaking}
\label{sec:bondbreaking}
The excitation conditions used in the previous section have been such that $\bra{\Psi(t)} H_{\rm mol} \ket{\Psi(t)}$ is well below the dissociation threshold of 20160 \cm{} (2.5 eV). Next, following previous work \cite{kuhn05_48}, the question will be explored whether the anharmonic coupling of the 4D PES suffices to cause bond dissociation at all. To this end the ground state wave packet will be displaced such that its center is at the positions on the PES given in Fig. \ref{fig:fig_pes}. These points have been chosen such that the expectation value of the energy is slightly above the dissociation energy, i.e. 20970 \cm{} (2.6 eV). This gives the initial state for a field-free wave packet propagation. The reaction yield is calculated according to Eq. \eqref{eq:yield}. 

\begin{figure}[t]
    \centering
        \includegraphics[width=0.9\textwidth]{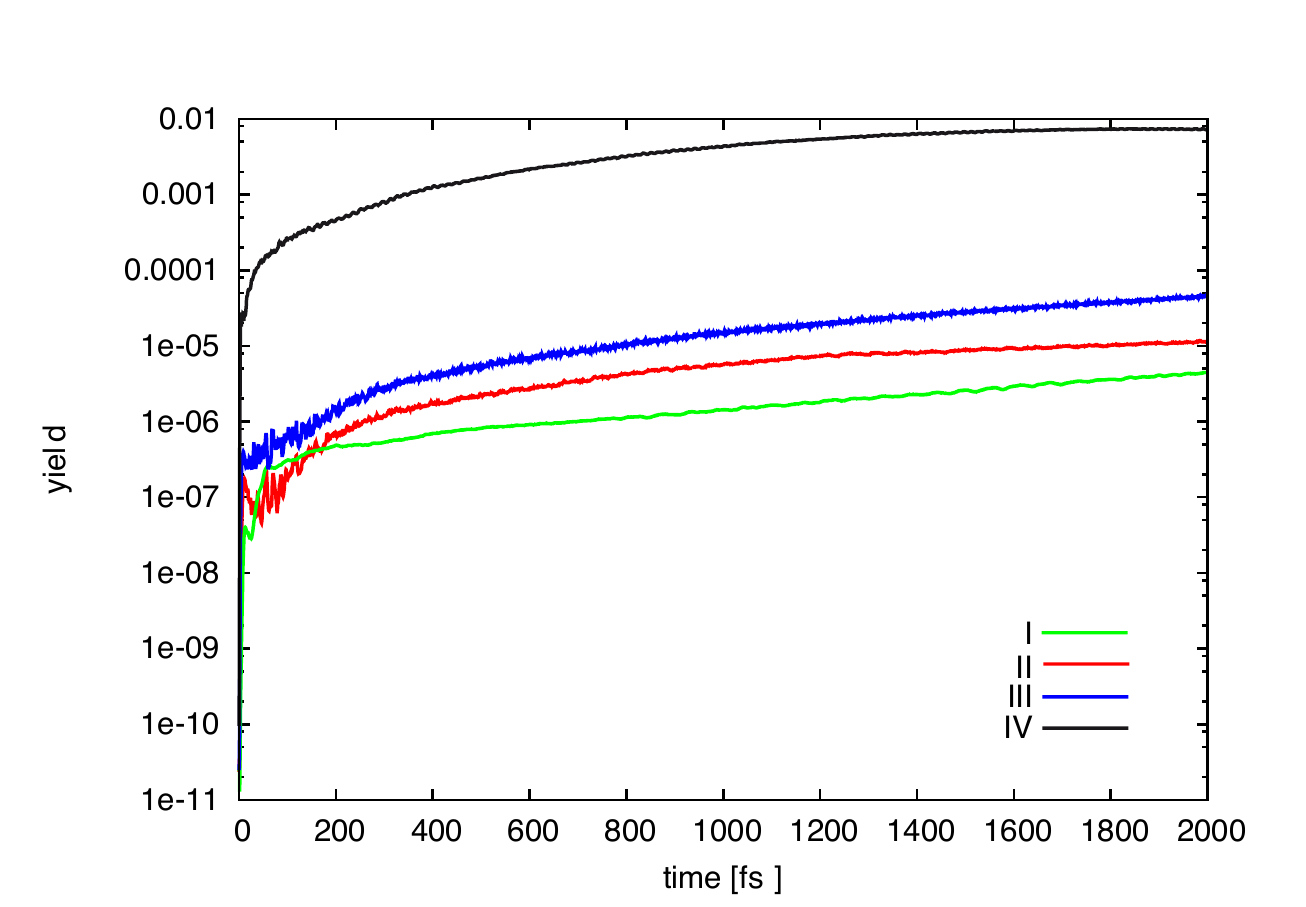}
    \caption{Reaction yield, Eq. \eqref{eq:yield}, for an initial state where the ground state wave packet is shifted to different positions on the PES as given in Fig. \ref{fig:fig_pes}. (I) $r_1=r_2=0.35$ a$ _{\rm B} $ (25 SPFs), (II) $r_1=0.55$ a$ _{\rm B} $ (10 SPFs), (III) $r_1=0.5$ a$ _{\rm B} $ and  $R_1=-0.1$ a$ _{\rm B} $ (12 SPFs), (IV) $R_1=-0.625$ a$ _{\rm B} $ (8 SPFs). The number of SPFs refers to the two combined modes and has been chosen such that the maximum of the smallest natural orbital population was below $\sim$0.1\% \cite{beck00_1}. 
    }
    \label{fig:yield}
\end{figure}

The results are summarized in Fig. \ref{fig:yield}. Apparently, the reaction yield is rather low on the time scale of 2 ps, i.e. similar to the case of the carboxymyoglobin model of Ref. \cite{kuhn05_48}. In particular, excitation of the carbonyl stretching coordinates, no matter whether it is collective (I) or localized (II) doesn't lead to a yield larger than $\sim$10$^{-5}$. Simultaneous displacement along the Co-CO bond coordinates (III) increases the yield by a factor of two. In fact, only an extreme compression of the Co-CO bond (IV) does give an appreciable reaction yield of about $\sim$10$^{-2}$. However, apart from the experimental constraints, driving the metal-carbonyl bond could be complicated by the fact, that in this spectral range modes that are not part of the present model would influence the dynamics, e.g., via IVR processes.

\section{Conclusions}
\label{sec:conc}
Previously, we had shown that collective CO vibrations in Mn$_2$(CO)$_{10}$ can be manipulated using circularly polarized laser pulses such that  the vibrational excitation circles around the molecular axis \cite{abdel-latif11_084314}. Here, we expanded on the topic of manipulating collective carbonyl vibrations and demonstrated the effect of localization of vibrational motion, achieved by two perpendicular linearly polarized laser pulses. The main results can be summarized as follows: (i) Given a pair of IR active symmetric and antisymmetric stretching vibrations, a superposition state can be excited, which corresponds to an alternating oscillation along a local bond stretching coordinate. The time  for switching between the two localized vibrations is determined by the inverse of the difference between the symmetric and antisymmetric transition frequency. (ii) By choosing the relative phase between the two overlapping pulses, the initially driven local coordinate can be determined. (iii) The efficiency of this process diminishes with increasing excitation level due to the anharmonicity leading to wave packet dispersion.

The simulations have been performed assuming that the molecule is fixed in space. In principle such condition could be realized, for instance,  by immobilization on a surface. In gas phase one could resort to intricate field-driven alignment and orientation schemes such as demonstrated in Ref. \cite{nevo09_9912}. In solution phase one can argue that the ultrashort overlapping pulses would preferentially excite those molecules, which have the proper orientation with respect to the laser polarization directions. Rotational motion during the pulses and even on the time scale considered here can certainly be neglected. Of course, when it comes to experimental observation care needs to be taken of the fact that actually a distribution of differently oriented molecules will be excited. 
Concerning the experimental signatures of the proposed transient localization we note that Ashihara et al \cite{ashihara13_05024} in their laser control experiment attributed an oscillatory signal in transient pump-probe spectroscopy to the creation of a coherent superposition between symmetric and antisymmetric fundamental transitions. How the present dynamics reflects in time-resolved nonlinear spectroscopy remains to be investigated.

Point (iii) above implies that this simple pulse scheme is not suitable for strong excitation of a local CO bond, such that IVR can cause dissociation of the neighboring metal-carbonyl bond. Whether more complicated pulse forms (see, e.g. Ref. \cite{zhao00_4882}) can achieve the goal of keeping the wave packet compact even at high energies remains to be shown. In any case, CpCo(CO)$_{2}$ appears to be another example where non-statistical metal-carbonyl bond breaking cannot be achieved by a reasonable IR excitation of the carbonyl stretching vibration.

\begin{acknowledgments}
The authors gratefully acknowledge financial support by the Deutsche Forschungsgemeinschaft (project Ku952/6).
\end{acknowledgments}

\begin{thebibliography}{10}

\bibitem{baiz09_1395}
C.~R. Baiz, P.~L. McRobbie, J.~M. Anna, E.~Geva, and K.~J. Kubarych.
\newblock Acc. Chem. Res. {\bf 42}, 1395 (2009).

\bibitem{baiz09_2433}
C.~R. Baiz, P.~L. McRobbie, N.~K. Preketes, K.~J. Kubarych, and E.~Geva.
\newblock J. Phys. Chem. A {\bf 113}, 9617 (2009).

\bibitem{rice01}
S.~Rice and M.~Zhao.
\newblock {\em {Optimal Control of Molecular Dynamics}\/} (Wiley, Hoboken,
  2001).

\bibitem{SfbBook}
O.~K{\"u}hn and L.~W{\"o}ste.
\newblock {\em {Analysis and Control of ultrafast Photoinduced Reactions}\/},
  vol.~87 of {\em Springer Series in Chemical Physics\/} (Springer, Heidelberg,
  2007).

\bibitem{shapiro11}
M.~Shapiro and P.~Brumer.
\newblock {\em {Principles of the Quantum Control of Molecular Processes}\/}
  (Wiley-VCH, Weinheim, 2011).

\bibitem{chelkowski90_2355}
S.~Chelkowski, A.~D. Bandrauk, and P.~B. Corkum.
\newblock Phys. Rev. Lett. {\bf 65}, 2355 (1990).

\bibitem{arrivo95_247}
S.~Arrivo, T.~Dougherty, W.~Grubbs, and E.~Heilweil.
\newblock Chem. Phys. Lett. {\bf 235}, 247 (1995).

\bibitem{maas98_75}
D.~J. Maas, D.~I. Duncan, R.~B. Vrijen, W.~J. v.~d. Zande, and L.~D. Noordam.
\newblock Chem. Phys. Lett. {\bf 290}, 75 (1998).

\bibitem{witte03_2021}
T.~Witte, T.~Hornung, L.~Windhorn, D.~Proch, R.~d. Vivie-Riedle, M.~Motzkus,
  and K.~L. Kompa.
\newblock J. Chem. Phys. {\bf 118}, 2021 (2003).

\bibitem{windhorn03_641}
L.~Windhorn, J.~S. Yeston, T.~Witte, W.~Fuss, M.~Motzkus, D.~Proch, K.-L.
  Kompa, and C.~B. Moore.
\newblock J. Chem. Phys. {\bf 119}, 641 (2003).

\bibitem{strasfeld09_1}
D.~B. Strasfeld, S.-H. Shim, and M.~T. Zanni.
\newblock Adv. Chem. Phys. {\bf 141}, 1 (2009).

\bibitem{strasfeld07_038102}
D.~B. Strasfeld, S.-H. Shim, and M.~T. Zanni.
\newblock Phys. Rev. Lett. {\bf 99}, 038102 (2007).

\bibitem{strasfeld09_105046}
D.~B. Strasfeld, C.~T. Middleton, and M.~T. Zanni.
\newblock New J. Phys. {\bf 11}, 105046 (2009).

\bibitem{ashihara13_05024}
S. Ashihara,  K. Enomoto, J. Tayama. 
\newblock EPJ Web of Conferences. \textbf{41,} 05024 (2013).

\bibitem{ventalon04_13216}
C.~Ventalon, J.~M. Fraser, M.~H. Vos, A.~Alexandrou, J.-L. Martin, and
  M.~Joffre.
\newblock Proc. Natl. Acad. Sci. USA {\bf 101}, 13216 (2004).

\bibitem{kuhn05_48}
O.~K\"uhn.
\newblock Chem. Phys. Lett. {\bf 402}, 48 (2005).

\bibitem{beck00_1}
M.~H. Beck, A.~J{\"a}ckle, G.~A. Worth, and H.-D. Meyer.
\newblock Phys. Rep. {\bf 324}, 1 (2000).

\bibitem{kuhn09_329}
O.~K\"uhn.
\newblock In {\em Multidimensional Quantum Dynamics\/} (edited by H.-D. Meyer,
  F.~Gatti, and G.~A. Worth) (Wiley-VCH, Weinheim, 2009), p. 329.

\bibitem{gollub07_369}
C.~Gollub, B.~M.~R. Korff, K.~L. Kompa, and R.~d. Vivie-Riedle.
\newblock PhysChemChemPhys {\bf 9}, 369 (2007).

\bibitem{meier05_044504}
C.~Meier and M.-C. Heitz.
\newblock J. Chem. Phys. {\bf 123}, 044504 (2005).

\bibitem{engel09_29}
V.~Engel, C.~Meier, and D.~J. Tannor.
\newblock Adv Chem Phys {\bf 141}, 29 (2009).

\bibitem{falvo13_145101}
C.~Falvo, A.~Debnath, and C.~Meier.
\newblock J. Chem. Phys. {\bf 138}, 145101 (2013).

\bibitem{debnath13_12884}
A.~Debnath, C.~Falvo, and C.~Meier.
\newblock J. Phys. Chem. A {\bf 117}, 12884 (2013).

\bibitem{orel00_94}
A.~E. Orel, Y.~Zhao, and O.~K\"{u}hn.
\newblock J. Chem. Phys. {\bf 112}, 94 (2000).

\bibitem{zhao99_7}
Y.~Zhao and O.~K\"{u}hn.
\newblock Chem. Phys. Lett. {\bf 302}, 7 (1999).

\bibitem{kuhn99_3103}
O.~K\"uhn, J.~Manz, and Y.~Zhao.
\newblock PhysChemChemPhys {\bf 1}, 3103 (1999).

\bibitem{zhao00_4882}
Y.~Zhao and O.~K\"uhn.
\newblock J. Phys. Chem. A {\bf 104}, 4882 (2000).

\bibitem{g09}
M.~J. Frisch, G.~W. Trucks, H.~B. Schlegel, G.~E. Scuseria, M.~A. Robb, J.~R.
  Cheeseman, G.~Scalmani, V.~Barone, B.~Mennucci, G.~A. Petersson,
  H.~Nakatsuji, M.~Caricato, X.~Li, H.~P. Hratchian, A.~F. Izmaylov, J.~Bloino,
  G.~Zheng, J.~L. Sonnenberg, M.~Hada, M.~Ehara, K.~Toyota, R.~Fukuda,
  J.~Hasegawa, M.~Ishida, T.~Nakajima, Y.~Honda, O.~Kitao, H.~Nakai, T.~Vreven,
  J.~J.~A. Montgomery, J.~E. Peralta, F.~Ogliaro, M.~Bearpark, J.~J. Heyd,
  E.~Brothers, K.~N. Kudin, V.~N. Staroverov, R.~Kobayashi, J.~Normand,
  K.~Raghavachari, A.~Rendell, J.~C. Burant, S.~S. Iyengar, J.~Tomasi,
  M.~Cossi, N.~Rega, J.~M. Millam, M.~Klene, J.~E. Knox, J.~B. Cross,
  V.~Bakken, C.~Adamo, J.~Jaramillo, R.~Gomperts, R.~E. Stratmann, O.~Yazyev,
  A.~J. Austin, R.~Cammi, C.~Pomelli, J.~W. Ochterski, R.~L. Martin,
  K.~Morokuma, V.~G. Zakrzewski, G.~A. Voth, P.~Salvador, J.~J. Dannenberg,
  S.~Dapprich, A.~D. Daniels, O.~Farkas, J.~B. Foresman, J.~V. Ortiz,
  J.~Cioslowski, and D.~J. Fox.
\newblock {Gaussian 09, Revision D.01,Wallingfort, CT} (2009).

\bibitem{dunietz03_5623}
B.~D. Dunietz, A.~Dreuw, and M.~Head-Gordon.
\newblock J. Phys. Chem. B {\bf 107}, 5623 (2003).

\bibitem{beagley79_47}
B.~Beagley, C.~T. Parrott, V.~Ulbrecht, and G.~G. Young.
\newblock J. Mol. Struct. {\bf 52}, 47 (1979).

\bibitem{meyer90_73}
H.-D. Meyer, U.~Manthe, and L.~S. Cederbaum.
\newblock Chem. Phys. Lett. {\bf 165}, 73 (1990).

\bibitem{mctdh84}
G.~Worth, M.~Beck, A.~J{\"a}ckle, and H.-D. Meyer.
\newblock {\em {The MCTDH Package, Version 8.2, (2000), University of
  Heidelberg, Heidelberg, Germany. H.-D. Meyer, Version 8.3 (2002), Version 8.4
  (2007), O. Vendrell and H.-D. Meyer, Version 8.5 (2011)}\/}.
\newblock {See http://mctdh.uni-hd.de} (2007).

\bibitem{jackle96_7974}
A.~J\"ackle and H.-D. Meyer.
\newblock J. Chem. Phys. {\bf 104}, 7974 (1996).

\bibitem{may11}
V.~May and O.~K\"uhn.
\newblock {\em {Charge and Energy Transfer Dynamics in Molecular Systems, 3rd
  Revised and Enlarged Edition}\/} (Wiley-VCH, Weinheim, 2011).

\bibitem{cotton55_175}
F.~A. Cotton, A.~D. Liehr, and G.~Wilkinson.
\newblock J. Inorg. Nucl. Chem. {\bf 1}, 175 (1955).

\bibitem{abdel-latif11_084314}
M.~Abdel-Latif and O.~K{\"u}hn.
\newblock J. Chem. Phys. {\bf 135}, 084314 (2011).

\bibitem{nevo09_9912}
I. Nevo, L. Holmegaard, J. H. Nielsen, J. L. Hansen, H. Stapelfeldt, F. Filsinger, G. Meijer, J. K\"upper.
\newblock PhysChemChemPhys. \textbf{11}, 9912 (2009).

\end{thebibliography}

\clearpage\newpage
\clearpage\newpage
\clearpage\newpage

\clearpage\newpage

\clearpage\newpage

\clearpage\newpage

\clearpage\newpage
\clearpage\newpage

\end{document}